\documentclass[aps,prd,reprint,superscriptaddress,nofootinbib,floatfix,longbibliography]{revtex4-1}
\usepackage{amssymb,amsmath,amsfonts,bm,slashed,soul,ragged2e,graphicx,epstopdf,hyperref,array,caption,subcaption}
\usepackage[utf8]{inputenc}
\usepackage[dvipsnames]{xcolor}
\usepackage[normalem]{ulem}
\usepackage[compat=1.1.0]{tikz-feynman}
\usepackage{soul}
\setstcolor{teal}
\hypersetup{colorlinks=true,urlcolor=blue,citecolor=blue,linkcolor=blue,menucolor=blue,anchorcolor=blue,filecolor=blue}
\enlargethispage{\baselineskip}
\DeclareCaptionJustification{justified}{\justifying}
\everymath{\displaystyle}
\renewcommand{\arraystretch}{1.5}

\newcommand{\mau}[1]{\textcolor{purple}{#1}}
\newcommand{\mg}[1]{\textcolor{purple}{{\bf [MG: #1]}}}
\newcommand{\tos}[1]{\textcolor{blue}{{\bf [TO'S: #1]}}}

\newcommand{\change}[1]{{#1}}
\begin{document}

\title{Solar chameleons: novel channels} 

\newcommand{\UniZar}{\affiliation{Departamento de F{\'i}sica Te{\'o}rica, Universidad de Zaragoza,\\C.\ de Pedro Cerbuna 12, 50009 Zaragoza, Spain}}
\newcommand{\CAPA}{\affiliation{Centro de Astropartículas y F{\'i}sica de Altas Energ{\'i}as (CAPA), Universidad de Zaragoza, C.\ de Pedro Cerbuna 12, 50009 Zaragoza, Spain}}

\author{Tom{\'a}s O’Shea}
\email{tgerard@unizar.es}
\UniZar\CAPA

\author{Anne-Christine Davis}
\email{ad107@cam.ac.uk}
\affiliation{Department of Applied Mathematics and Theoretical Physics (DAMTP),\\Center for Mathematical Sciences, University of Cambridge, CB3 0WA, United Kingdom}
\affiliation{Kavli Institute for Cosmology (KICC), University of Cambridge,\\Madingley Road, Cambridge CB3 0HA, United Kingdom}

\author{Maurizio Giannotti}
\email{mgiannotti@unizar.es}
\UniZar\CAPA

\author{Sunny Vagnozzi}
\email{sunny.vagnozzi@unitn.it}
\affiliation{Department of Physics, University of Trento, Via Sommarive 14, 38123 Povo (TN), Italy}
\affiliation{Trento Institute for Fundamental Physics and Applications (TIFPA)-INFN,\\Via Sommarive 14, 38123 Povo (TN), Italy}

\author{Luca Visinelli}
\email{luca.visinelli@sjtu.edu.cn}
\affiliation{Tsung-Dao Lee Institute (TDLI), No.\ 1 Lisuo Road, 201210 Shanghai, China}
\affiliation{School of Physics and Astronomy, Shanghai Jiao Tong University,\\Dongchuan Road 800, 200240 Shanghai, China}

\author{Julia K. Vogel}
\email{julia.vogel@cern.ch}
\UniZar\CAPA

\date{\today}

\begin{abstract}

\noindent We revisit the flux of chameleons (light scalar particles which could play a role in the dark energy phenomenon) produced in the interior of the Sun. 
Our novel analysis incorporates various important details and new processes that have previously been overlooked, including the impact of the bulk magnetic field profile, as well as Primakoff production of chameleons in the electric fields of electrons and ions.
In this paper we consider only the contributions of transverse photons. The production of chameleons from longitudinal electromagnetic excitations will be presented in a dedicated follow-up work.
Demanding that the total flux of chameleons does not exceed 3\% of the solar luminosity leads to the stringent upper limit on the chameleon-photon conformal coupling $\beta_\gamma \lesssim 10^{10}$, assuming that the height of the chameleon potential is set to the dark energy scale $\Lambda = 2.4$\,meV, and independently of other couplings to matter. Although this bound is tighter than current upper limits on $\beta_{\gamma}$ from the CAST helioscope, these limits will have to be reassessed in terms of the updated solar chameleon flux we have computed. We argue that solar chameleons, potentially detectable in next-generation helioscopes such as IAXO, can be used to probe a region of chameleon parameter space that has yet to be covered.
\end{abstract}

\maketitle

\section{Introduction}
\label{sec:introduction}

The search for new light (typically bosonic) particles is a steadily growing endeavor which continues to spark interest and intense activity within the experimental, theoretical, and phenomenology communities~\cite{Irastorza:2018dyq,DiLuzio:2020wdo,Lanfranchi:2020crw,Mitridate:2022tnv,Antel:2023hkf}. The reason is, at the very least, three-fold. Firstly, light particles appear ubiquitously in some of the most well-motivated extensions of the Standard Model of particle physics (SM), including but not limited to string theory~\cite{Svrcek:2006yi,Arvanitaki:2009fg,Cicoli:2012sz,Visinelli:2018utg,Cicoli:2023opf}, as well as bottom-up extensions of the SM where otherwise accidental symmetries are gauged~\cite{Carone:1994aa,Davidson:1978pm,Mohapatra:1980qe,Davidson:1987mh,Foot:1990mn,He:1991qd,FileviezPerez:2010gw}. Furthermore, important technological advances have placed a wide range of experimental facilities, now reaching maturity or soon to start taking data, in the position of being able to probe or exclude a number of key theoretical benchmark scenarios, whereas important complementary constraints are expected at the same time from cosmological and astrophysical observations (see e.g.\ Refs.~\cite{CAST:2008ixs,Ehret:2010mh,Graham:2012su,Borexino:2012guz,Ringwald:2012hr,Budker:2013hfa,Essig:2013goa,Foot:2014uba,Hlozek:2014lca,Arvanitaki:2014wva,Baumann:2015rya,Caldwell:2016dcw,Perivolaropoulos:2016ucs,Poulin:2018dzj,Davoudiasl:2019nlo,Nojiri:2019riz,Vagnozzi:2019ezj,KumarPoddar:2020kdz,Hagstotz:2020ukm,Odintsov:2020iui,Croon:2020oga,Rogers:2020ltq,Giare:2020vzo,Oikonomou:2020qah,Caputo:2020sys,Caputo:2021efm,Tsai:2021irw,Xu:2021rwg,Calza:2021czr,Roy:2021uye,Vagnozzi:2022moj,DEramo:2022nvb,Poulin:2023lkg,Tsai:2023zza,Oikonomou:2023qfl}). Last but definitely not least, new light degrees of freedom may be at the origin of the dark matter (DM) and dark energy (DE) components which add up to 95\% of the Universe's energy budget~\cite{Hui:2016ltb,Ferreira:2020fam,Choi:2021aze}.

The accelerated cosmological expansion first inferred in 1998~\cite{Riess:1998cb,Perlmutter:1998np} and now corroborated by a wide variety of probes~\cite{Rubin:2016iqe,Nadathur:2020kvq,DiValentino:2020evt} is often explained assuming that around 70\% of the energy budget of the Universe is in the form of an exotic component referred to as dark energy, whose nature however remains a mystery to date~\cite{Nojiri:2006ri,Frieman:2008sn,Bamba:2012cp,Nojiri:2017ncd,Huterer:2017buf}. The simplest possibility is one where DE is related to a cosmological constant $\Lambda_{\rm DE} \approx 2.4\,$meV -- however, if the microphysical origin thereof is interpreted in terms of zero-point vacuum energy density of quantum fields~\cite{Carroll:2000fy,Shapiro:2003kv,Shapiro:2006qx,Sola:2013gha,Padilla:2015aaa}, such a scenario suffers from a number of severe fine-tuning problems~\cite{Weinberg:1988cp} (see, however, Refs.~\cite{Moreno-Pulido:2022phq,SolaPeracaula:2022hpd}). Another possibility of particular interest resolves around the idea that the microphysics of DE resides in a new light boson yet to be discovered. Such a ``quintessence'' field would be most easily distinguishable from a cosmological constant through its time-varying imprint on the cosmological background~\cite{Ratra:1987rm,Wetterich:1987fm,Caldwell:1997ii,Zlatev:1998tr,Linder:2007wa}. Intriguingly, recent results from the Dark Energy Spectroscopic Instrument (DESI) in combination with other cosmological probes tentatively hint towards a strong preference for an evolving DE component~\cite{DESI:2024mwx,Calderon:2024uwn,DESI:2024kob} (see also Refs.~\cite{Tada:2024znt,Yin:2024hba,Luongo:2024fww,Wang:2024hks,Cortes:2024lgw,Colgain:2024xqj,Carloni:2024zpl,Berghaus:2024kra,Gomez-Valent:2024tdb,Wang:2024dka,Yang:2024kdo,Park:2024jns,Huang:2024qno,Dinda:2024kjf,Bousis:2024rnb}), potentially constituting the signature of a quintessence field~\cite{Shlivko:2024llw}. Novel signatures of quintessence fields, and more generally evolving DE components, are also among the key science goals of ongoing and next-generation cosmological surveys~\cite{Spergel:2013tha,Amendola:2016saw,CMB-S4:2016ple,SimonsObservatory:2018koc,SimonsObservatory:2019qwx}.

A scenario where a new light, possibly (pseudo)scalar, degree of freedom may be at the origin of the DE phenomenon and thereby explain cosmic acceleration is a particularly intriguing one~\cite{Sahni:2002kh}. However, such a scenario also poses significant challenges. Firstly, in most models this degree of freedom would have to be extremely light, at least on cosmological scales, in order to be frozen or nearly frozen by Hubble friction. Additionally, unless a protection mechanism or symmetry is at play, such a particle would naturally couple to matter fields, typically at least with gravitational strength~\cite{Carroll:1998zi,Amendola:1999er}. This would give rise to unobserved, and thereby undesired, long-range fifth forces~\cite{Dvali:2001dd,Upadhye:2006vi,Upadhye:2012qu}, alongside a host of other potentially interesting effects which can be searched for on astrophysical and cosmological scales~\cite{Wetterich:2002ic,Uzan:2002vq,Vagnozzi:2019kvw,BeltranJimenez:2020iyx,Trojanowski:2020xza,Ferlito:2022mok,Hoshiya:2022ady,Archidiacono:2022iuu,Kaneta:2023wdr,Jimenez:2024bhv}. However, unless one is willing to fine-tune the model parameters, a mechanism which allows for significant deviations from General Relativity on cosmological scales, while dynamically suppressing fifth forces on local scales (for instance, altering the force's strength, range, or behaviour as a function of distance), is required in order for such a scenario to be phenomenologically viable. Many such \textit{screening} mechanisms have been proposed. Examples include the chameleon~\cite{Khoury:2003rn,Khoury:2003aq}, symmetron~\cite{Hinterbichler:2010es}, environmental dilaton~\cite{Damour:1994zq}, and Vainshtein~\cite{Vainshtein:1972sx} mechanisms: we refer the reader to Ref.~\cite{Brax:2021wcv} for a recent review on screening mechanisms.

In this work, our focus is going to be on chameleon-screened scalars, wherein the field acquires a density-dependent \textit{effective} mass $m_{\text{eff}}$. Other cases of scalar particles following in a similar fashion are beyond the scope of this work. In high-density environments -- such as on local scales, where the existence of fifth forces is much more constrained -- the chameleon's mass $m_{\text{eff}}$ becomes large, leading in turn to a short-ranged fifth force which allows chameleons to escape detection in solar system or terrestrial searches for fifth forces~\cite{Khoury:2003aq,Khoury:2003rn,Brax:2004qh,Burrage:2016bwy,Burrage:2017qrf}. On the other hand, chameleons can propagate freely in low-density environments, for instance on cosmological scales. Such a density-dependent behaviour is made possible by virtue of a direct coupling of the scalar field to the local density (as a result, such a scenario can also be interpreted in terms of the scalar field leading to a modification of gravity).

Intriguingly, theories equipped with screening mechanisms are amenable to local tests of their associated effects, opening the window towards terrestrial, and more generally local, tests of DE (see e.g.\ Refs.~\cite{Brax:2007ak,Brax:2007vm,GammeV:2008cqp,Brax:2009ey,Steffen:2010ze,Baum:2014rka,Lemmel:2015kwa,Elder:2016yxm,Brax:2016did,Sakstein:2019qgn,Desmond:2019ygn,Pernot-Borras:2021edr,Cai:2021wgv,Katsuragawa:2021wmw,Benisty:2021cmq,Brax:2021owd,Yuan:2022cpw,Brax:2022olf,Benisty:2022lox,Hogas:2023vim,Elder:2023oar,Benisty:2023dkn,Benisty:2023vbz,Briddon:2023ayq,Hogas:2023pjz,Benisty:2023clf,Kumar:2024ylj} for a rich array of examples in the chameleon case). As alluded to earlier, despite their elusive nature, a rich array of experiments is aimed at detecting new light bosons, for instance through direct detection in liquid Xenon~\cite{XENON:2022ltv,LZ:2023poo,Wang:2023wrr}, Argon~\cite{DarkSide:2022knj}, resonant conversion in cavities~\cite{ADMX:2010ndb,Alesini:2023qed}, photon-chameleon-photon or ``afterglow'' transitions with laser experiments~\cite{Ahlers:2007st,Gies:2007su}, and opto-mechanical force sensors~\cite{ArguedasCuendis:2019fxj} among others. It is therefore natural to ask whether at least part of the experimental setup currently in place can be used to detect (chameleon-)screened light particles, which could ultimately be linked to the DE problem.

One environment which has attracted particular attention in this context is the Sun. The reason why the Sun is among the favorite laboratories for particle physicists is to be sought in its hot and dense environment, its proximity compared to other astrophysical objects, as well as the presence of a strong magnetic field, all of which conspire to provide a natural setting for the potential production of new light particles. Chameleons can be produced in the Sun through a variety of mechanisms. One proposed channel involves the conversion of thermal photons into chameleons in the Sun's magnetic field~\cite{Raffelt:1987im}. If sufficiently weakly coupled, these scalar bosons could then escape the Sun, contributing to its energy loss. This argument can be used to constrain the chameleon couplings in much the same way as with axions, dark photons, and so forth. Furthermore, chameleons escaping the Sun could reach the Earth, leading to the possibility of direct detection thereof. Indeed, solar (but more generally stellar) production is often the starting point for studies which look into the possibility of detecting new light particles~\cite{Raffelt:1996wa} (see e.g.\ Refs.~\cite{Raffelt:1990yz,Paschos:1993yf,Moriyama:1995bz,Redondo:2013lna,Giannotti:2015kwo,Giannotti:2017hny,Chang:2018rso,Budnik:2019olh,Carenza:2020zil,OHare:2020wum,An:2020bxd,Lucente:2020whw,Pallathadka:2020vwu,Dev:2020jkh,Carenza:2020cis,Edwards:2020afl,Caputo:2021kcv,Caputo:2021eaa,Calore:2021klc,Fischer:2021jfm,Caputo:2021rux,Berlin:2021kcm,Caputo:2022mah,Lucente:2022vuo,Balaji:2022noj,Fiorillo:2022cdq,Bottaro:2023gep,Lella:2023bfb,Hoof:2023jol,Carenza:2023lci} for examples of studies in this direction).

From the considerations laid out above, the question of whether the terrestrial experimental setup in place is suitable to detect chameleon-screened particles is somewhat intertwined with the issue of the production of chameleons in the Sun. Surprisingly, this is a question which has received very little attention to date. Indeed, the state-of-the-art for what concerns production of solar chameleons dates back to two (it is fair to say not widely known) works from 2010 and 2011 respectively~\cite{Brax:2010xq,Brax:2011wp}. The problem was reconsidered in 2021 by some of the authors of the present work (together with one of the authors of Refs.~\cite{Brax:2010xq,Brax:2011wp}) in Ref.~\cite{Vagnozzi:2021quy}. However, the update was limited to the detection channel. Specifically, Ref.~\cite{Vagnozzi:2021quy} studied the possibility that interactions between solar chameleons and electrons in direct detection chambers may have been responsible for the excess observed by the XENON1T experiment~\cite{XENON:2020rca}, albeit with no significant changes with respect to the original work. However, the model adopted to study solar chameleon production in these works is rather simplistic, envisaging exclusively Primakoff-like production in the solar magnetic field, with the latter approximated as being a narrow shell located in the tachocline. All of these are clearly approximations at best. To draw a parallelism, the issues of production of solar axions~\cite{Redondo:2013wwa,Wu:2024fsf,Caputo:2024oqc} and subsequent terrestrial detection thereof~\cite{Dimopoulos:1986mi,Pospelov:2008jk,Li:2015tsa} (and similarly for hidden photons as well as other light bosons) is instead one which has received significantly more attention, despite the underlying mathematical and physical tools being admittedly rather similar.

Regardless of the reason behind the lag in the solar chameleon literature relative to the solar axion and hidden photon cases, the time is ripe to revisit the issue of production of chameleons in the Sun, and fill in a number of gaps present in the state-of-the-art. This is therefore the goal of the present work. We provide a thorough investigation of various points which were overlooked in previous literature. For instance, we extend the Primakoff production mechanism to include also the production of chameleons in the electric fields of electrons and ions present in the solar plasma. 
This is not a mere repetition of what has already been done for axions, as it requires a careful treatment of longitudinal photon modes (plasmons) which, due to selection rules, do not play a role in the solar axion case. A detailed study of the plasmon processes will be covered in a dedicated follow-up work (part II of this series). Moreover, we consider the full solar magnetic field profile, rather than limiting the latter to a thin shell around the tachocline, and discuss in detail solar luminosity bounds on chameleons, in particular on their couplings to photons and matter. Our results lead us to identify significant differences with respect to the previous state-of-the-art, and grant us a much better understanding of the resulting spectrum of solar chameleons in the ${\cal O}({\text{eV}}$--${\text{keV}})$ range, while also opening the window towards the possibility of experimentally distinguishing solar chameleons from solar axions. Our investigations should therefore ultimately be understood as being preparatory to a study on the detectability of solar chameleons, and possibly the physics of DE, in terrestrial laboratories. We plan to address these very important issues in a follow-up paper.

The rest of this work is then organized as follows. In Sec.~\ref{sec:methods} we discuss in detail the methods and assumptions adopted in our calculation of the solar chameleon spectrum. The results of this calculation are presented in Sec.~\ref{sec:results}, which presents a detailed overview of the various processes we consider. We critically discuss our findings in Sec.~\ref{sec:discussion}, where we provide a detailed comparison to other existing experimental setups. Finally, in Sec.~\ref{sec:conclusions} we draw concluding remarks and outline interesting directions for possible follow-up work. A detailed derivation of the photon-chameleon conversion probability is provided in Appendix~\ref{sec:appendix}. The code used for the numerical analysis carried out in this work is made available on \texttt{GitHub} at \url{https://github.com/tomasoshea/chameleon}. Throughout our work we make use of natural units with $\hbar=c=1$, unless otherwise indicated.

\section{Methods}
\label{sec:methods}

\subsection{Modeling the solar interior}

Assessing the composition of the Sun at its different layers plays a significant role for both astrophysical models of stellar evolution and the understanding of fundamental physics such as neutrino oscillations~\cite{SNO:2001kpb}. For this, the inner distribution of the Sun is a subject of intense work that has led to the definition of a \emph{standard solar model}, see e.g.\ Refs.~\cite{Bahcall:1995bt,Christensen-Dalsgaard:1996hpz,Gough:1996am}. Models for the solar interior include but are not limited to GN93~\cite{prantzosorigin},
GS98~\cite{Grevesse:1998bj}, AGSS09~\cite{Asplund:2009fu}, C11~\cite{Caffau:2010qc}, AGSS15~\cite{Scott:2014lka,Scott:2014mka}, B16~\cite{Vinyoles:2016djt}, AAG21~\cite{2021A&A...653A.141A}, and B23~\cite{herrera_2024_10822316}.

Here, we adopt the AGSS09 solar model~\cite{Asplund:2009fu}, which is based on a description of the Sun as a spherically symmetric and quasi-static star. Although this is not the most recent model in the literature, it is well suited for our purpose and allows us to directly compare our results with previous studies in which AGSS09 is also used -- we do not expect significant changes to our results were we to use the latest solar model.\footnote{\change{A detailed analysis of the uncertainties due to different solar models for the axion case can be found in Ref.~\cite{Hoof:2021mld}. The uncertainty in the Primakoff flux is estimated to be around 10-15\%. We expect a similar uncertainty for the chameleon case, though we have not performed an explicit analysis with the different models.}}
Within the adopted model, the stellar structure is specified by a set of differential equations and boundary conditions for the luminosity, radius, age and composition of the Sun~\cite{Bahcall:2005va}.\footnote{The code is archived at \href{https://wwwmpa.mpa-garching.mpg.de/~aldos/solar\_main.html}{https://wwwmpa.mpa-garching.mpg.de/{\string~}aldos/solar\_main.html}. See in addition the model by Bahcall presented in Ref.~\cite{Bahcall:2004fg} and archived at \href{http://www.sns.ias.edu/~jnb/SNdata/Export/BP2004/bp2004stdmodel.dat}{http://www.sns.ias.edu/bp2004stdmodel.dat}.}

Though in the standard solar models discussed in the previous references  the Sun is modeled as a quasi-static environment, there exist seismic solar models which include large-scale magnetic fields in different regions of the solar interior~\cite{Turck-Chieze:2001aug,Couvidat:2002gvk}. According to these models, the \textit{radiative} zone ($r \lesssim 0.7 \; R_\odot$) hosts a magnetic field of intensity $B_\text{rad} \in [200 \,{\rm T},\,3000 \,{\rm T}]$; the \textit{tachocline} ($ r \sim 0.7 \; R_\odot$) a magnetic field of intensity $B_\text{tach} \in [4 \,{\rm T},\,50 \,{\rm T}]$; and the exterior (\textit{convective}) zone ($r \gtrsim 0.9 \; R_\odot$) a magnetic field of intensity $B_\text{conv} \in [3 \,{\rm T},\,4 \,{\rm T}]$~\cite{Hoof:2021mld}. So far, only the contribution from the tachocline has been considered in the solar chameleon literature. We shall see that the core magnetic field provides a larger contribution to the resulting flux. \change{See Sec.~\ref{sec:discussion} for a discussion of the validity of using a static solar composition and magnetic field model.}

The magnetic field is modeled by assuming three distinct sections of quadrupolar solar magnetic fields, with profiles given by the following~\cite{Couvidat:2002gvk}:
\begin{equation}
    {\bf B}(r, \vartheta) = 3B_i(r) \cos(\vartheta) \sin(\vartheta) \hat{\bf e}_\phi\,,
\end{equation}
where $r$ is the radial coordinate, $\vartheta$ the polar angle, and the functional form of $B_i$ is:
\begin{equation}
    B_i(r) = K_i\,\left(\frac{r}{d}\right)^{\alpha_i}\,\left(1 - \left(\frac{r-r_1}{d}\right)^2\right)^{\beta_i}\,\hat{B}_i\,.
    \label{eq:Bfieldsolarmodel}
\end{equation}
The fitting parameters entering into Eq.~(\ref{eq:Bfieldsolarmodel}) are listed in Tab.~\ref{table:Bfieldmodel}, with the index $i$ labeling the different zones.

\begin{table}
    \begin{center}
    \renewcommand{\arraystretch}{1.4}
    \begin{tabular}{|@{\hspace{0.1 cm}}l@{\hspace{0.1 cm}}|@{\hspace{0.1 cm}}l@{\hspace{0.1 cm}}|@{\hspace{0.1 cm}}l@{\hspace{0.1 cm}}|@{\hspace{0.1 cm}}l@{\hspace{0.1 cm}}|}
    \hline
    Parameter & Radiative zone & Tachocline & Outer layers\\
    \hline\hline
    $\hat{B}_i$ & \change{200 - 3000}\,T & \change{4 - 50}\,T & \change{3 - 4}\,T\\
    $K_i$ & $(1+\lambda)(1+1/\lambda)^\lambda$ & 1 & 1\\
    $\alpha_i$ & 2 & 0 & 0\\
    $\beta_i$ & $\lambda$ & 1 & 1\\
    $d$ & $0.712\,R_\odot$ & 0.02\,$R_\odot$ & 0.035\,$R_\odot$\\
    $r_1$ & 0 & $0.732\,R_\odot$ & 0.96\,$R_\odot$\\
    \hline
    \end{tabular}
    \end{center}
    \caption{Coefficients for the solar magnetic field model in Eq.~\eqref{eq:Bfieldsolarmodel}, taken from Ref.~\cite{Couvidat:2002gvk}. Some parameters are defined in terms of the tachocline radius $r_0 = 0.712\,R_\odot$ and the parameter $\lambda = 1+10\,r_0/R_\odot$.}
    \label{table:Bfieldmodel}
\end{table}

\begin{figure}[!ht]
    \centering
    \includegraphics[width=\linewidth]{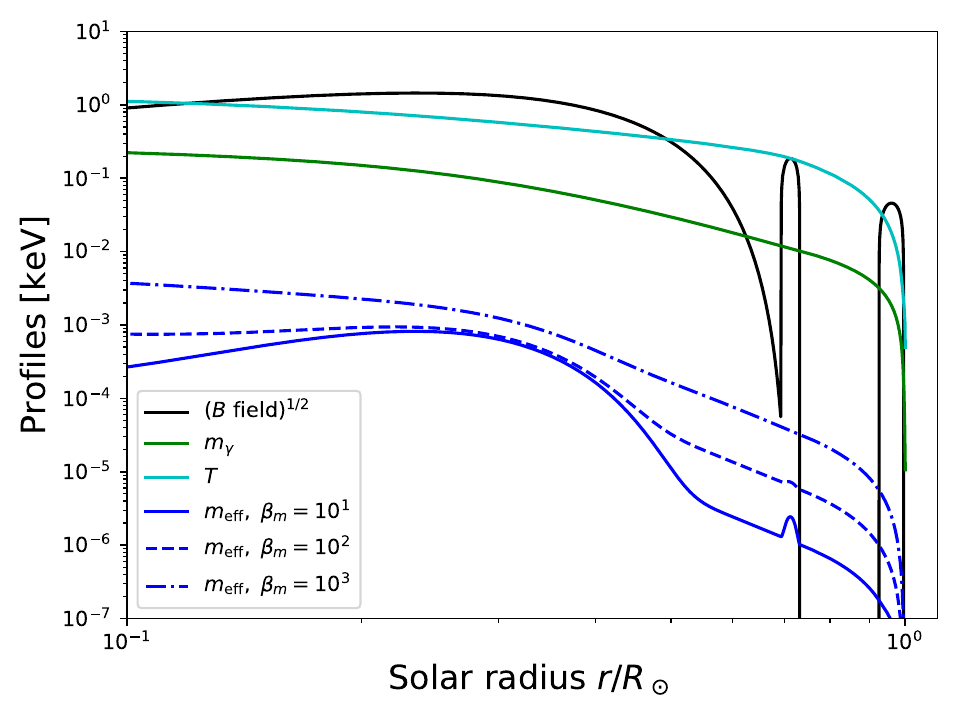}
    \caption{Radial profiles (as a function of distance from the center of the Sun, in units of solar radius) for the square root of the magnetic field in the Sun (black curve), the plasma frequency/effective photon mass (green curve), the Sun's temperature profile (cyan curve), and the chameleon effective mass $m_{\text{eff}}$ (blue curves), with the latter calculated assuming $n=1$, chameleon-photon coupling $\beta_\gamma=10^{10}$, and different values of the chameleon-matter coupling $\beta_m=10^1$ (solid curve), $10^2$ (dashed curve), and $10^3$ (dot-dashed curve).
    \change{The magnetic field shown is the upper limit given in Tab.~\ref{table:Bfieldmodel}.}}
    \label{fig:Bfield}
\end{figure}

The solar model provides the profiles for the temperature $T$ and the density $\rho$ as a function of the inner radius $r$. Assuming charge neutrality, the electron number density in terms of the atomic mass unit $m_u$ and the mean molecular weight per electron $\mu_e$ is then $n_e = \rho/(m_u \mu_e)$. The corresponding plasma-induced effective photon mass can be expressed as follows~\cite{Raffelt:1996wa}:
\begin{equation}
    m_\gamma = \left(\frac{4\pi \alpha_{\text{em}} n_e}{m_e}\right)^{1/2}\,,
    \label{eq:omegap}
\end{equation}
with $\alpha_{\text{em}} \approx 1/137.036$ being the fine structure constant. Fig.~\ref{fig:Bfield} reports the square root of the magnetic field profile with the parameters of Tab.~\ref{table:Bfieldmodel} (black curves), along with the profiles for the temperature (cyan curve) and the plasma frequency given by Eq.~\eqref{eq:omegap} (green curve). Also shown is the effective mass of the chameleon (blue curves), discussed later in the text and given by Eq.~(\ref{eq:m_eff}). The profile of the latter is given for the parameters of the chameleon theory $n=1$, the chameleon-photon coupling $\beta_\gamma=10^{10}$, the height of the chameleon potential appearing in Eq.~(\ref{eq:potential}) set to the ``DE scale'' $\Lambda = \Lambda_{\rm DE}=2.4\,\text{meV}$, and for different values of the chameleon-matter coupling: $\beta_m = 10^1$ (solid blue curve), $\beta_m = 10^2$ (dashed blue curve), and $\beta_m = 10^3$ (dot-dashed blue curve). The parameters $\beta_\gamma$ and $\beta_m$ characterize the strength of the interactions of the chameleon with photons and matter fields respectively, and their meaning is discussed later in the text, in Sec.~\ref{sec:Photon-chameleon_system}.

\subsection{Photon-chameleon system}
\label{sec:Photon-chameleon_system}

We consider a theory featuring a chameleon field $\phi$ modeled as a real scalar field and described by the following action~\cite{Brax:2010uq}:
\begin{equation}
	\label{eq:action}
    \begin{split}
    S =&{} \int {\rm d}^4x\,\sqrt{-g}\,\mathcal{L}_E(g_{\mu\nu},\phi) \\
    & + \int {\rm d}^4x\,\sqrt{-\tilde g}\,\sum_i\,\mathcal{L}_m(\tilde g_{\mu\nu}, \psi_i) + S_{\rm SM}\,.
    \end{split}
\end{equation}
In the above action, the first term describes the dynamics of the scalar field coupled to the metric tensor $g_{\mu\nu}$ in the Einstein frame, and is given by the following:
\begin{equation}
	\mathcal{L}_E(g_{\mu\nu},\phi) = X - V_{\rm self}(\phi)\,,
\end{equation}
where $X = -(1/2)(\partial_\mu\phi)(\partial^\mu\phi)$ is the kinetic term of the chameleon field, and $V_{\rm self}(\phi)$ is the self-interacting potential. As per standard choice in the field, we parametrize it through an energy scale $\Lambda$ and an index $n$ as follows~\cite{Ratra:1987rm}:
\begin{equation}
    \label{eq:potential}
    V_{\rm self}(\phi) = \Lambda^4 \left(1 + \frac{\Lambda^n}{\phi^n}\right)\,.
\end{equation}
In the literature, the quantity $\Lambda$ for which cosmic acceleration in agreement with observations is recovered is often fixed to the value $\Lambda_{\rm DE} = 2.4\,$meV, usually referred to as the ``DE scale''. The value used for the quantity $\Lambda_{\rm DE}$ serves as a benchmark over which different experimental results are compared.

The second term in Eq.~\eqref{eq:action} describes the coupling of the chameleon to the matter fields $\psi_i$ for the $i$th species through a universal coupling to the metric in the Jordan frame $\tilde g_{\mu\nu}$. As a consequence of diffeomorphism invariance of GR, the relation between the metrics in the Einstein and Jordan frames can generically include functions of both the scalar field $\phi$ and its kinetic term, in the following form~\cite{Bekenstein:1992pj,Zumalacarregui:2010wj,Sebastiani:2016ras}:
\begin{equation}
\tilde g_{\mu\nu} = \mathcal{A}^2(\phi, X)\,g_{\mu\nu} + \mathcal{B}^2(\phi, X)\,\partial_\mu\phi\partial_\nu\phi\,,
\end{equation}
where the functions $\mathcal{A}(\phi, X)$ and $\mathcal{B}(\phi, X)$ describe the conformal and disformal components of the transformation respectively, and in general can be functions of both $\phi$ and its kinetic term $X$. In what follows we make the following choice for consistency with Ref.~\cite{Vagnozzi:2021quy}:
\begin{eqnarray}
    \label{eq:a}
    \mathcal{A}^2(\phi, X) &=& 1+2\beta_i\frac{\phi}{M_{\rm Pl}}\,,\\
    \mathcal{B}^2(\phi, X) &=& \frac{2}{M_i^4}\,.
    \label{eq:b}
\end{eqnarray}
In particular, with our choice neither $\mathcal{A}$ nor $\mathcal{B}$ depend on $X$ (this implies, for instance, that we do not include a kinetic-conformal coupling -- see Ref.~\cite{Vagnozzi:2021quy} for more in-depth discussions on the rationale behind the above choice). In Eqs.~(\ref{eq:a},\ref{eq:b}), $\beta_i$ are the species-specific conformal couplings and the energy scale $M_i$ parametrizes the strength of the disformal coupling of the chameleon with the $i$th species. The chameleon couples to matter and radiation fields through the coupling to the metric in the ``Jordan frame'' $\tilde g_{\mu\nu}$, which acts as an effective dimension-five interaction that practically introduces an effective density-dependent coupling. Finally, the last term in Eq.~\eqref{eq:action} denotes the SM action:
\begin{equation}
	S_{\rm SM} = \int {\rm d}^4x\,\sqrt{-g}\,\left(\mathcal{L}_{\rm EM} + \mathcal{L}_m\right)\,,
\end{equation}
where the interaction of the electromagnetic (EM) field $A_\mu$ with the electric current $J^\mu$ is described by the Maxwell Lagrangian:
\begin{equation}
    \mathcal{L}_{\rm EM} = -\frac{1}{4}F^{\mu\nu}F_{\mu\nu} + A_\mu J^\mu\,,
\end{equation}
with the EM field strength being $F^{\mu\nu} = \partial^\mu A^\nu - \partial^\nu A^\mu$, while $\mathcal{L}_m$ describes the kinetic motion of the plasma field inside the Sun.

In terms of the effective theory under consideration, the action in Eq.~\eqref{eq:action} can be expressed as follows:
\begin{equation}
	\label{eq:action1}
	S = \int\! {\rm d}^4x\sqrt{-g}\,\left[ X \!-\! V_\mathrm{eff}(\phi) \!+\! \frac{1}{M_\gamma^4}(\partial_\mu\phi)(\partial_\nu\phi)T_\gamma^{\mu\nu} \!\right] \!+ S_{\rm SM},
\end{equation}
where $M_\gamma$ is the energy scale for the disformal coupling associated to the chameleon interacting with the photon field, whose stress-energy tensor is given by the following:
\begin{equation}
T^{\mu\nu}_\gamma = F^{\mu\alpha}{F^\nu}_\alpha - \frac{1}{4}\,g^{\mu\nu}\,F^{\alpha\beta}F_{\alpha\beta}\,.
\end{equation}
Finally, the effective potential to which the chameleon field responds combines the self-interacting potential and the conformal couplings, and is given by the following:
\begin{equation}
    V_\mathrm{eff}(\phi) = V_\mathrm{self}(\phi) + \frac{\beta_m}{M_\mathrm{Pl}}\rho_m\phi + \frac{\beta_\gamma}{M_{\rm Pl}}\phi \frac{1}{4}F^{\mu\nu}F_{\mu\nu}\,,
    \label{eq:V_eff}
\end{equation}
where $\beta_\gamma$ is the coupling constant for the conformal chameleon-photon interaction, $\rho_m$ is the energy density of non-relativistic matter at the relevant location inside the Sun, and $\beta_m$ is an effective coupling to non-relativistic matter that results from modeling the interaction with the dense matter environment in the solar interior in terms of the energy density of the $i$th field $\rho_i$ as follows:
\begin{equation}
    \sum_i \beta_i\rho_i = \beta_m\rho_m\,.
\end{equation}
The action in Eq.~\eqref{eq:action1} governing the evolution of the chameleon field leads to the equations of motion provided in Appendix~\ref{sec:appendix}. In what follows, we neglect the disformal coupling in Eq.~\eqref{eq:action1} since the energy scale $M_\gamma$ could be several orders of magnitude above the current bounds $M_\gamma \gtrsim 10^{-3}\,$eV~\cite{Brax:2013nsa} (see also Ref.~\cite{Vagnozzi:2021quy} for further discussions on this point).

Although neither the self-interacting potential given in Eq.~\eqref{eq:potential} nor the interactions individually feature a minimum, the effective potential $V_\mathrm{eff}(\phi)$ does present such a minimum for a field value $\phi_{\rm min}$ in the presence of an ambient matter and electromagnetic energy density. For example, assuming the presence of a bulk magnetic field $B$ in the solar environment leads to the condition:
\begin{equation}
    -\frac{\partial V_{\rm self}}{\partial \phi}(\phi_{\rm min}) = \frac{\beta_m}{M_\mathrm{Pl}}\rho_m + \frac{\beta_\gamma}{2 M_\mathrm{Pl}} B^2\,.
\end{equation}
As discussed further in Sec.~\ref{sec:discussion}, 
even the largest allowed values of the solar magnetic field provide a negligible impact in the above equation with respect to the contribution from $\rho_m$, so that we can safely approximate the density-dependent minimum as follows:
\begin{equation}
    \phi_{\rm min} \approx \left( \frac{n \Lambda^{n+4} M_\mathrm{Pl}}{\beta_m \rho_m} \right)^\frac{1}{n+1}\,.
    \label{eq:phi0}
\end{equation}
Accordingly, the corresponding density-dependent effective mass in an environment with matter density $\rho_m$ is given by the following:
\begin{equation}
    m^2_{\text{eff}}(\rho_m) = \frac{\partial^2 V_{\rm self}}{\partial \phi^2}(\phi_{\rm min}) \approx n(n+1)\Lambda^{n+4} \left( \frac{\beta_m \rho_m}{n M_\mathrm{Pl} \Lambda^{n+4}} \right)^{\frac{n+2}{n+1}}\,.
    \label{eq:m_eff}
\end{equation}

\section{Results}
\label{sec:results}

In what follows, we will numerically determine the solar chameleon flux. We begin by reconsidering the production of chameleons in the solar magnetic field, considering also the contribution from the core magnetic field (whereas previously only the tachocline had been considered). Additionally, we consider Primakoff production in the electric fields of ions and electrons in the Sun which has been overlooked in the previous literature. Our derivations assume a scalar field featuring a conformal coupling to photons and a mass term $\mathcal{L} \supset -\frac{1}{2}m^2\phi^2$ -- therefore, our results can be applied to a fixed-mass scalar, or alternatively to a chameleon field upon replacement of $m$ with the density-dependent effective mass given in Eq.~\eqref{eq:m_eff}.

\subsection{Chameleon production in bulk magnetic fields}
\label{sec:tft}

\begin{figure}
\centering
\begin{tikzpicture}
    \begin{feynman}
        \vertex (a) {\(\phi\)};
        \vertex [right=of a] (b);
        \vertex [right=of b] (c);
        \vertex [right=of c] (d) {\(\phi\)};
        \vertex [below=of b, crossed dot] (B1) {\(\)};
        \vertex [below=of c, crossed dot] (B2) {\(\)};

        \diagram* {
            (a) -- [scalar] (b) -- [photon, edge label = \(\gamma\)] (c) -- [scalar] (d),
            (b) -- [photon, edge label = \(\mathbf{B}\)] (B1),
            (c) -- [photon, edge label = \(\mathbf{B}\)] (B2),
        };
    \end{feynman}
\end{tikzpicture}
\caption{Feynman diagram relevant for the lowest-order contribution to the self-energy of scalars $\phi$ arising from the scalar-photon coupling in the presence of an external B-field.}
\label{fig:feynman_TFT}
\end{figure}
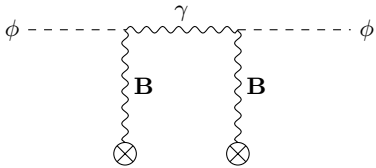

The thermal field theory approach is one of the most straightforward and elegant ways of estimating the chameleon production rate in the solar magnetic field. For the axion case, this is discussed, for example, in Ref.~\cite{Caputo:2020quz} and in appendix C of Ref.~\cite{Guarini:2020hps}, which we closely follow here.

The Lagrangian in Eq.~\eqref{eq:action1} contains the following term:
\begin{equation}
    \mathcal{L} \supset \frac{\beta_\gamma}{M_\mathrm{Pl}} \mathbf{B} \cdot (\nabla \phi \times \mathbf{A})\,.
    \label{eq:lagrangian_TFT}
\end{equation}
Note that we ignore the disformal coupling since, as argued in Ref.~\cite{Vagnozzi:2021quy}, the disformal channel must be subdominant in the magnetic field Primakoff process due to energy loss considerations in horizontal branch stars. The diagram shown in Fig.~\ref{fig:feynman_TFT} describes the following lowest-order contribution to the scalar self-energy in a plasma:
\begin{equation}
    \pi_\phi = \left( \frac{\beta_\gamma B_\perp k_\phi}{M_\mathrm{Pl}} \right)^2 \frac{(\omega_\gamma^2 - k_\gamma^2 - \pi_\gamma)^*}{\vert\omega_\gamma^2 - k_\gamma^2 - \pi_\gamma \vert^2}\,,
    \label{eq:pi_phi}
\end{equation}
where $B_\perp$ is the component of the B-field perpendicular to $\mathbf{k}$, and $\pi_\gamma$ is the photon self-energy, which takes the following form:
\begin{equation}
    \pi_\gamma \approx m_\gamma^2 - i \omega \Gamma_\gamma\,,
    \label{eq:pi_gamma}
\end{equation}
with the real part $m_\gamma$ being the plasma frequency, interpreted as the effective mass of the transverse photon modes, given by Eq.~\eqref{eq:omegap}. $\Gamma_\gamma$ is instead related to the rates for photon production, $\Gamma_\gamma^{\mathrm{pro}}$, and absorption, $\Gamma_\gamma^{\mathrm{abs}}$, by the general relation~\cite{Weldon:1983jn}:
\begin{equation}
    \Gamma = \Gamma^{\mathrm{abs}} - \Gamma^{\mathrm{pro}} = \left(1-e^{-\omega/T} \right) \Gamma^{\mathrm{abs}} = \left( e^{\omega/T} - 1 \right) \Gamma^{\mathrm{pro}}\,.
\end{equation}
We see that the rate of production of scalars is related to the self-energy by the following relation:
\begin{equation}
    \Gamma_\phi^{\mathrm{pro}} = \frac{-\mathrm{Im}[\pi_\phi]}{\omega(e^{\omega/T} - 1)}\,,
    \label{eq:Gamma_phi--weldon}
\end{equation}
which combined with Eq.~\eqref{eq:pi_phi} leads to:
\begin{equation}
    \Gamma_\phi^{\mathrm{pro}} = \left( \frac{\beta_\gamma B_\perp}{M_\mathrm{Pl}} \right)^2 \frac{k_\phi^2}{\vert\omega^2 - k_\gamma^2 - \pi_\gamma \vert^2} \frac{\Gamma_\gamma}{e^{\omega/T} -1}\,.
    \label{eq:Gamma_phi}
\end{equation}
Note that the sum over transverse photon polarizations is implicitly included in the definition of $B_\perp^2 = B_x^2 + B_y^2$, where $x$ and $y$ are the directions orthogonal to the direction of propagation. The function $\Gamma_\gamma(\omega,r)$ accounts for the production and absorption of photons in the Sun by various processes: it is discussed at length in Ref.~\cite{Redondo:2015iea} and summarized in the Appendix of Ref.~\cite{OShea:2023gqn}.

In this work we assume that only the Thomson and free-free contributions are important, an assumption which is valid for high energies. The form of $\Gamma_{\gamma}$ used here is shown explicitly in Eq.~\eqref{eq:Gamma_gamma}. The reason for neglecting additional contributions to $\Gamma_\gamma$ requires further clarification. In fact, these terms could in principle be relevant at low energies. For example, it was shown that the full form of $\Gamma_\gamma$ adds a significant flux in the case of low energy solar hidden photons, for energies $\omega \lesssim 10\,$eV~\cite{OShea:2023gqn}, and in this work we are interested in chameleons with energies down to $1\,$eV. However, in our case the additional terms can be ignored as the chameleon field at such low energies depends on the weak magnetic field at the solar surface. Furthermore, preliminary studies into the plasmon contributions appear to show that the magnetic field contribution is sub-dominant at all energies. Therefore, it can be shown that the effect of including extra photon absorption processes has a negligible impact on the total production rate of solar chameleons.

Finally, to obtain the total production rate $\dot N$ we need to integrate over the phase space, as follows:
\begin{equation}
    \mathrm{d} \dot N = \mathrm{d}V \frac{\mathrm{d}^3k_\phi}{(2\pi)^3} \Gamma_\phi^{\operatorname{prod}}\,.
    \label{eq:dN}
\end{equation}
This can be written as:
\begin{equation}
    \frac{\mathrm{d}\dot N}{\mathrm{d}\omega} = \frac{2 \beta_\gamma^2}{\pi M_\mathrm{Pl}^2} \int_0^{R_\odot} \mathrm{d}r\,r^2 B_\perp^2(r) \frac{\omega(\omega^2 - m^2)^{3/2}}{|\omega^2 - k_\gamma^2 - \pi_\gamma|^2} \frac{\Gamma_\gamma}{e^{\omega/T} - 1}\,.
    \label{eq:dN_TFT--1}
\end{equation}
If we consider the case $p_\gamma = p_\phi$, the above simplifies to the following:
\begin{equation}
    \frac{\mathrm{d}\dot N}{\mathrm{d}\omega} \!=\! \frac{2 \beta_\gamma^2}{\pi M_\mathrm{Pl}^2} \!\int_0^{R_\odot} \!\!\mathrm{d}r r^2 B_\perp^2(r) \frac{\omega(\omega^2 - m^2)^{3/2}}{(m_\gamma^2 - m^2)^2 \!+\! (\omega \Gamma_\gamma)^2} \frac{\Gamma_\gamma}{e^{\omega/T} - 1}\,.
    \label{eq:dN_TFT--2}
\end{equation}
Alternatively, the production rate can be derived through a kinetic approach, as shown in Appendix~\ref{sec:appendix}.

\subsection{Primakoff production from charged particles}
\label{sec:primakoff}

\begin{figure}
    \centering
\begin{tikzpicture}
    \begin{feynman}
        \vertex (a1);
        \vertex [right=of a1] (a11);
        \vertex [right=of a11] (b1);
        \vertex [right=of b1] (b11);
        \vertex [right=of b11] (c1);
        \vertex [below=of a1] (a2);
        \vertex [below= of b1] (b2);
        \vertex [below=of c1] (c2);

        \diagram* {
            (a1) -- [photon, edge label = \(\gamma (p_\gamma)\)] (b1) -- [scalar, edge label =  \(\phi (p_\phi)\)] (c1),
            (b1) -- [photon, edge label = \(\gamma (p_q)\)] (b2),
            (a2) -- [fermion, edge label =  \(e^- (p_1)\)] (b2) -- [fermion, edge label =  \(e^- (p_2)\)] (c2),
        };
    \end{feynman}
\end{tikzpicture}
    \caption{Feynman diagram relevant for Primakoff production of scalars $\phi$ from electron-photon interactions.}
    \label{fig:feynmanElectron}
\end{figure}
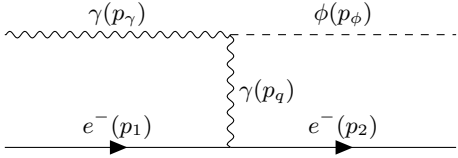

The Lagrangian terms relevant for Primakoff production are the following:
\begin{equation}
    \mathcal{L} \supset \frac{-\beta_\gamma}{4M_{\rm Pl}} \phi F_{\mu\nu}F^{\mu\nu} - e A_\mu \bar\psi \gamma^\mu \psi \,.
    \label{eq:lagrangian_interaction}
\end{equation}
These allow for production of scalars through the interactions of photons with charged particles (electrons and ions) in the solar plasma, $\gamma+Z e \to Z e+\phi$. This process has been overlooked in the past. We find, however, that it contributes significantly to the emission rate of solar chameleons. Furthermore, the flux generated through this process suffers only from marginal astrophysical uncertainties, since it depends mostly on the solar temperature, density and chemical composition, which are known quite well. On the other hand, the process of chameleon production in the solar macroscopic magnetic field, studied in the previous section, is relatively more uncertain due to the less well known structure of the magnetic field inside the Sun (especially in the radiative zone). Note that here we have again ignored the disformal coupling, leaving the study of this production channel for future works.


The diagram relevant for Primakoff production through interactions with electrons is shown in Fig.~\ref{fig:feynmanElectron} (the diagram for production through interactions with ions is completely analogous), and its matrix element takes the following form:
\begin{equation}
    \mathcal{M}_\lambda = \left(\bar u_2 \gamma_\mu u_1 \right) \Gamma_\lambda^\mu\,,
    \label{eq:M}
\end{equation}
where $\lambda$ stands for the photon polarization state, $u_i$ is the spinor for electron $i$ with four-momentum $p_i$ and spin state $s_i$, and $\Gamma_\lambda^\mu$ contains all the non-spinor parts of the matrix element, given by the following:
\begin{equation}
    \Gamma_\lambda^\mu \equiv \frac{-\beta_\gamma e}{M_\mathrm{Pl} p_q^2} \left[ (p_q \cdot p_\gamma) \epsilon_\lambda^\mu - (p_q \cdot \epsilon_\lambda) p_\gamma^\mu \right]\,.
    \label{eq:G_lambda}
\end{equation}
In the above, we have used $1/p_q^2$ as the photon propagator, ignoring medium-induced effects as these will be included later through a screening factor. In the limit $m_e \to \infty$, the spin-averaged sum of the matrix element squared for any $\mathcal{M}$ of the form given by Eq.~\eqref{eq:M} reduces to the following:
\begin{equation}
    \left| \bar{\mathcal{M}} \right|_\lambda^2 \to \left| 2 m_e \Gamma_\lambda^0 \right|^2\,,
\end{equation}
assuming the initial and final electron spins are the same, which can be assumed in the case of scalars. In this paper we will consider only the transverse contributions, but there also exists a significant contribution at low energies from longitudinal plasmons. These will be considered in detail in a separate paper (part II of this analysis).

\begin{figure}
    \centering
    \includegraphics[width=\linewidth,keepaspectratio]{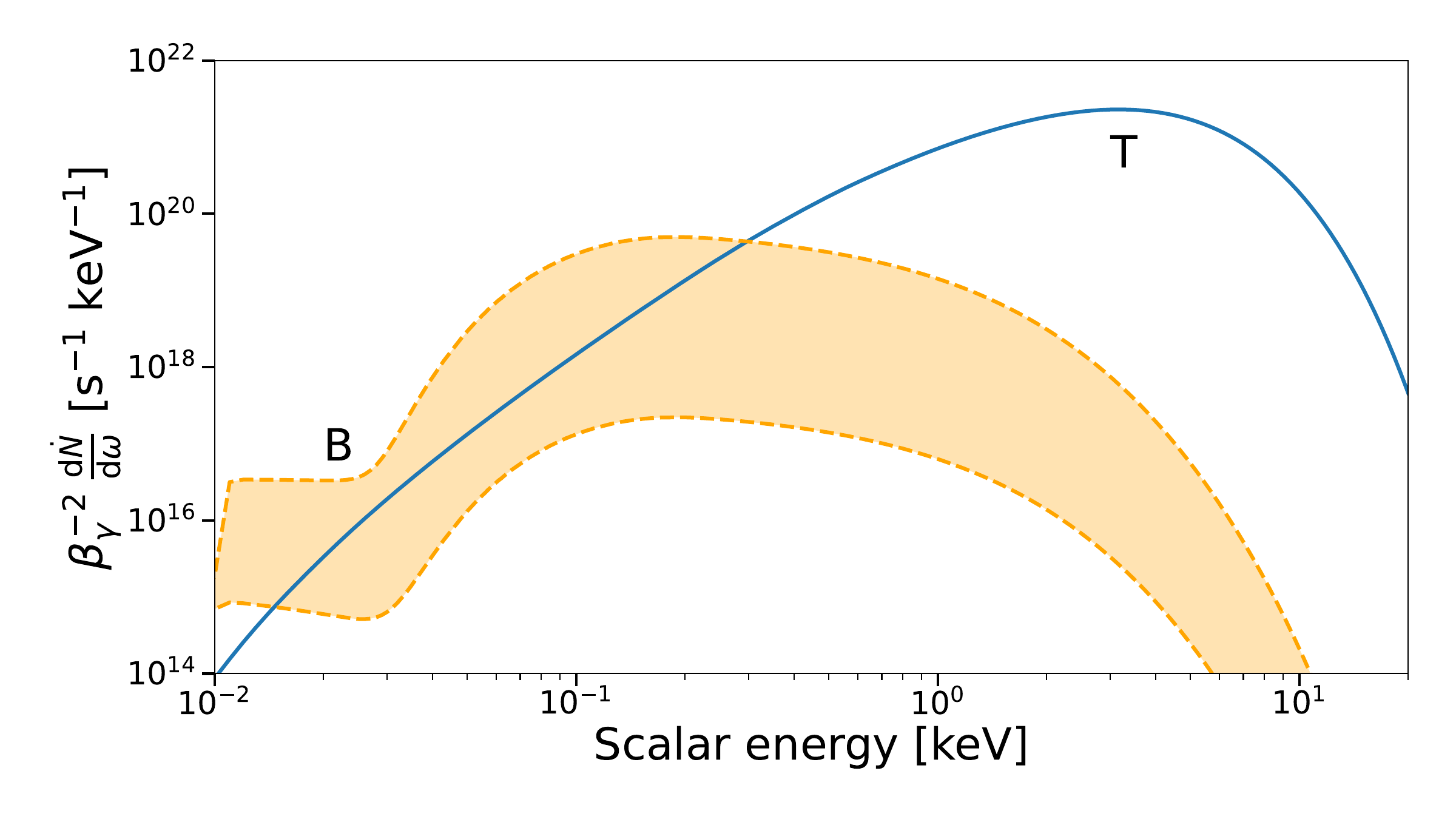}
    \caption{Comparison of the emission spectra (differential particle production rate per unit energy) of solar chameleons arising via Primakoff production from (transverse) photons (``T'', solid teal curve) in the presence of charged particles, as well as from production in the bulk magnetic field (``B'', dot-dashed orange curve). \change{The width of the orange band reflects the uncertainty on the value of the magnetic field strength. See Sec.~\ref{sec:methods} for details.} Note that the production rate is normalized by $\beta_{\gamma}^2$, in order to factor out the dependency on the chameleon-photon coupling. We have adopted the AGSS09 stellar model~\cite{Asplund:2009fu}, whereas the chameleon parameters are fixed to $\beta_m = 10^2$ and $n=1$, with the height of the potential set to the dark energy scale $\Lambda_{\text{DE}} = 2.4$\,meV.
    \change{Note that the low-energy contribution from the outer convective zone is not displayed due to the very low production rate.}}
    \label{fig:primakoff-LT}
\end{figure}

Averaging over the 2 transverse photon polarisations, we get the following contribution to the matrix element:
\begin{equation}
    \left| \bar{\mathcal{M}} \right|_{\rm t}^2 = \frac{2 \beta_\gamma^2 e^2 m_e^2}{M_\mathrm{Pl}^2} \frac{k_\phi^2 \omega_\gamma^2}{q^4} (1 - x^2)\,,
    \label{eq:M_t}
\end{equation}
where $e$ is the electron charge, $m_e$ is the electron mass, $p_q = (E_q, \mathbf{q}) = p_\phi - p_\gamma = p_1 - p_2$ is the 4-momentum transferred, and $x \equiv \cos(\bf{k_\phi}, \bf{k_\gamma})$. For the initial and final particles we have used the notation $p_a \equiv (\omega_a, \mathbf{k}_a)$. We have also used the fact that in the $m_e \to \infty$ limit, $p_q^2 \to -q^2$. It has been shown~\cite{Raffelt:1985nk,Raffelt:1987yu,Raffelt:1996wa} that the photon propagator in the non-degenerate plasma should be corrected to account for Debye screening by replacing $1/q^2 \to S(q)/q^2$, where $S(q)$ takes the following form:
\begin{equation}
  S(q) \equiv \frac{q^2}{q^2 + \kappa^2}\,,
  \label{eq:raffelt-screening}
\end{equation}
and $\kappa$ is the Debye screening scale:
\begin{equation}
    \kappa^2 = \frac{4 \pi \alpha}{T} \sum_i Z_i^2 n_i\,,
    \label{eq:debye}
\end{equation}
with the sum extending over all charged particles present in the Sun, and $n_i$ being the number density of particle $i$. The matrix element in Eq.~(\ref{eq:M_t}) therefore takes the following form:
\begin{equation}
    \left\vert \bar{\mathcal{M}} \right\vert_{\rm t}^2 = \frac{2 \beta_\gamma^2 e^2 m_e^2}{M_\mathrm{Pl}^2} k_\phi^2 \omega_\gamma^2 \frac{1 - x^2}{q^2(\kappa^2 + q^2)}\,.
    \label{eq:M_t--screened}
\end{equation}
Using the above form for $\vert\bar{\mathcal{M}}\vert^2_\mathrm{t}$, we can now compute the total production rate of scalars from this process, which takes the following form:
\begin{equation}
    \frac{\mathrm{d}\dot N}{\mathrm{d}\omega} = \frac{\beta_\gamma^2 \alpha}{8\pi M_\mathrm{Pl}^2} \int_0^{R_\odot} \frac{r^2 \mathrm{d}r}{e^{\omega/T} - 1} \frac{\omega^2 k_\phi}{k_\gamma} \mathcal{I}(u,v) \sum_i Z_i^2 n_i \,,
    \label{eq:production_t}
\end{equation}
where we have summed over all charged particles in the Sun, and $\mathcal{I}(u,v)$ is given by the following integral:
\begin{equation}
    \begin{split}
    &\mathcal{I}(u,v) \equiv \int_{-1}^{+1}  \frac{1 - x^2}{(u - x)(v + u - x)} {\rm d}x\\
    &= \frac{(u+v)^2 -1}{v} \ln \left( \frac{u+v+1}{u+v-1} \right) \!-\! \frac{u^2-1}{v} \ln \left( \frac{u+1}{u-1} \right) \!-\! 2\,.
    \end{split}
    \label{eq:I(uv)}
\end{equation}
Finally, in Eq.~\eqref{eq:production_t} we have defined the quantities $u$ and $v$ as follows:
\begin{eqnarray}
  &&  u = \frac{k_\gamma}{2 k_\phi} + \frac{k_\phi}{2 k_\gamma}\,, 
  \label{eq:u} \\
   && v = \frac{\kappa^2}{2 k_\gamma k_\phi}\,.
   \label{eq:v}
\end{eqnarray}
Making use of the dispersion relations $k_\gamma = \sqrt{\omega^2 - m_\gamma^2}$ and $k_\phi = \sqrt{\omega^2 - m^2}$, we can then calculate the production rate of either fixed-mass scalars, or chameleon-screened scalars. A comparison between the contributions to the solar chameleon flux resulting from the transverse photons interacting with the electric fields of electrons and ions (solid teal curve) and the bulk magnetic field channel discussed earlier (dot-dashed orange curve), is shown in Fig.~\ref{fig:primakoff-LT}. These are normalized by $\beta_{\gamma}^2$, in order to factor out the dependency on the chameleon-photon coupling, given that these fluxes scale as $\beta_{\gamma}^2$.

\subsection{Photon coalescence}
\label{sec:coalescence}

Aside from the processes considered so far, the scalar-photon interaction potentially allows for other processes to play a role in the production of solar chameleons. In particular, these are the photon coalescence ($\gamma \gamma \to \phi$) and plasmon decay ($\gamma_T \to \gamma_L \phi$) processes, which should in principle be allowed as production processes for $\phi$. In the case of axions ($a$), it was shown that the production rates from the processes $\gamma_L \gamma_T \to a$ and $\gamma_T \to \gamma_L a$ reduce to the $e^- \gamma_T \to e^- a$ production rate in the limit $\kappa \to \infty$, implying that there is no new contribution from these processes~\cite{Raffelt:1987np,Raffelt:1996wa}.
The processes involving plasmons can be shown to be limiting cases of the Primakoff process~\cite{Raffelt:1987np,Raffelt:1996wa,Liang:2023jlz}, as will be discussed further in part II of the series.
This implies the only truly new process is the coalescence of two transverse photons $\gamma_T \gamma_T \to \phi$. However, such a channel is subject to the kinematical restriction $m_{\text{eff}}^2 \geq 2 m_{\gamma}^2$, and for the range of parameters in which we are interested we have explicitly checked that $m_{\text{eff}} \ll m_\gamma$ holds everywhere in the Sun. This implies that solar production of scalars does not enjoy contributions from photon coalescence or plasmon decay processes, and that all of the production through the 2-photon vertex comes from the Primakoff process.

\section{Discussion}
\label{sec:discussion}

As a light particle, the chameleon is subject to a wide range of bounds arising from fifth force searches. Various tests for the existence of a fifth force have been performed to search for chameleons and other elusive particles, with no evidence for such fifth forces so far. These bounds translate into limits on the parameters that specify the chameleon model, namely $\beta_m$, $\beta_\gamma$, $n$, and $\Lambda$. We now critically discuss how our results can be related to the bounds obtained from such searches.

One powerful technique devised to search for chameleons is atom interferometry~\cite{Burrage:2014oza,Fischer:2024eic} which, setting the height of the potential to the DE scale $\Lambda = \Lambda_{\rm DE}$, provides the bound $\beta_m \lesssim 10^3$ for $n \lesssim 10$~\cite{Hamilton:2015zga,Jaffe:2016fsh,Sabulsky:2018jma}. In particular, for $n=1$, the 95\% confidence level (C.L.) upper limit $\beta_m \lesssim 360$ is obtained. For the same theoretical setup, torsion balance experiments instead set $\beta_m \gtrsim 10$. For $n=1$, this leaves a window in parameter space, $10 \lesssim \beta_m \lesssim 10^3$, that can be covered via experiments testing for violations of the equivalence principle, such as the E\"{o}t-Wash~\cite{Kapner:2006si,Upadhye:2012qu,Wagner:2012ui} and MICROSCOPE experiments~\cite{MICROSCOPE:2019jix,MICROSCOPE:2022doy}. In fact, the allowed window for $n=1$ chameleons corresponding to a cosmologically viable DE region has recently been probed with levitated force sensors~\cite{Yin:2022geb}, excluding the region $5 \lesssim \beta_m \lesssim 630$ at 95\% CL.

Aside from laboratory bounds, searches for solar chameleons can place an independent bound on the chameleon-photon coupling $\beta_\gamma$. Such a search performed in the CAST experiment initially resulted in the 95\%~C.L. upper limit $\beta_\gamma \lesssim 10^{11}$~\cite{CAST:2015npk}, later refined to $\beta_\gamma \lesssim 5.7\times 10^{10}$~\cite{CAST:2018bce}. Here, we demand that the luminosity carried away by solar chameleons $L_\phi$ resulting from our computation does not exceed 3\% of the total solar luminosity $L_\odot$, as inferred from recent solar studies performing global fits to helioseismic and solar neutrino observables~\cite{Vinyoles:2015aba}. These considerations result in the following upper limit on the chameleon-photon coupling:
\begin{equation}
    \beta_\gamma \lesssim 10^{10}\,.
    \label{eq:solarluminosity}
\end{equation}
While the above limit we obtained from energy loss considerations is slightly more stringent than the limit reported by the CAST collaboration, the latter analysis only accounts for the production of chameleons in the tachocline, in the same vein as Ref.~\cite{Vagnozzi:2021quy}. Therefore, a new analysis accounting for the new production channels and therefore flux contributions presented here should be carried out in order to assess the correct bound CAST can place on $\beta_{\gamma}$. Even without explicitly performing such an analysis, we can expect that the resulting constraint should actually be more stringent than the $\beta_\gamma \lesssim 5.7\times 10^{10}$ limit reported previously, since the flux resulting from production of chameleons in the tachocline provides a lower limit to the total flux accounting for all the channels we discussed here.
\change{Although a reliable calculation of the bound will require more sophisticated limit setting techniques involving CAST data and the complete chameleon spectrum including longitudinal contributions, a preliminary analysis suggests that the updated CAST bound would be around $\beta_\gamma \lesssim 6\times10^8$, a significant improvement even with respect to the solar energy loss bound set in this paper. A detailed study, including all contributions and experimental data is in preparation.}

\begin{figure}
    \centering
    \includegraphics[width=\linewidth]{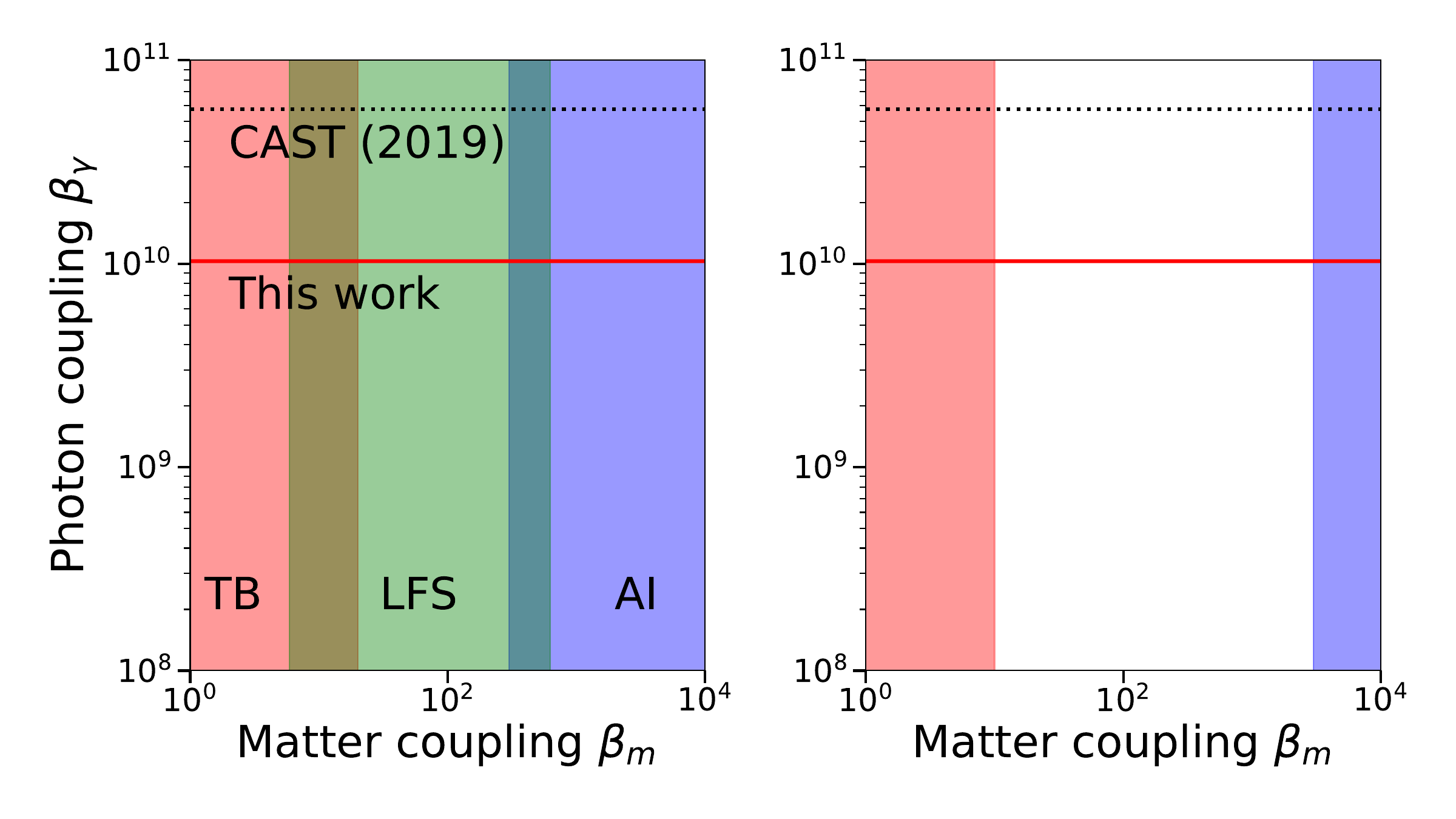}
    \caption[l]{95\% confidence level upper limit on the chameleon-photon coupling $\beta_{\gamma}$ obtained from demanding that the luminosity carried by solar chameleons does not exceed 3\% of the total solar luminosity (red horizontal line, with the region above the line excluded), while fixing $\Lambda$ to the dark energy scale $\Lambda_{\text{DE}}=2.4\,{\text{meV}}$, assuming $n=1$ (left panel) and $n=4$ (right panel). We also show the upper limit from CAST~\cite{CAST:2018bce} (black horizontal line, again with the region above the line excluded), whereas the vertical shaded bands indicated regions of the chameleon-matter coupling $\beta_m$ excluded from atom interferometry (``AI'', blue region), levitated force sensors (``LFS'', green region, only for $n=1$ chameleons in the left panel), and torsion balance experiments (``TB'', red region). Note that the CAST bound was calculated assuming only production of solar chameleons from the magnetic field in the tachocline: recalculating the bound adopting the full spectrum we have studied in the present work is expected to lead to a tighter limit. \change{See Sec.~\ref{sec:discussion} for more details.} }
    \label{fig:luminosity}
\end{figure}

Focusing on $n=1$ chameleons (while still setting $\Lambda = \Lambda_{\rm DE}$), the laboratory and solar bounds translate into the constraints on the couplings $\beta_m$ and $\beta_{\gamma}$ shown in Fig.~\ref{fig:luminosity} (left panel), where we display the three bounds on $\beta_m$ arising from atom interferometry (blue region), torsion balance experiments (red region), and levitated force sensors (green region), along with the upper bounds on $\beta_\gamma$ obtained from the CAST experiment (black horizontal line) and from our analysis (red horizontal line). Given that the regions enclosed within the bands are those excluded by the corresponding experiments, we see that levitated force sensors in principle have closed the remaining viable window of parameter space for $n=1$ chameleons, as discussed in Ref.~\cite{Yin:2022geb}.

The situation is different if we change the potential index $n$. In this case, the levitated force sensor constraints of Ref.~\cite{Yin:2022geb} do not apply, as these have been derived assuming $n=1$. On the other hand, the results derived in the present work, and thereby our constraint reported in Eq.~(\ref{eq:solarluminosity}), are virtually independent of $n$ as long as it is not too large ($n \lesssim 200$, with extremely large values of $n$ being anyhow of limited theoretical interest). The right panel of Fig.~\ref{fig:luminosity} is therefore analogous to the left panel discussed previously, but focusing on $n=4$ chameleons, for which the atom interferometry and torsion balance constraints weaken slightly. In this case, laboratory searches do not completely close the available $\beta_m$ parameter space, but leave the window $10 \lesssim \beta_m \lesssim 3\times 10^3$ open. We also note that, for general $n$, the laboratory bounds can be weakened by a slight reduction in the value of the cosmological parameter $\Lambda$: in particular, the levitated force sensor constraints can be evaded for $\Lambda \lesssim 1$\,meV. Obviously, reducing $\Lambda$ implies that some degree of fine-tuning of an overall additive constant is required if the chameleon particle in question is to drive cosmic acceleration.\footnote{Such tuning is obviously also present in self-accelerating scenarios, defined as acceleration in the Jordan frame but not the Einstein frame in the complete absence of any cosmological constant. In this case, the latter is therefore arbitrarily set to zero, thus leaving the so-called ``old cosmological constant'' problem open. In passing, we note that a powerful no-go theorem excludes the possibility of chameleon self-acceleration~\cite{Wang:2012kj}, but leaves open the possibility of a quintessence chameleon driving cosmic acceleration (see e.g.\ Ref.~\cite{Martin:2008qp}).}

A few comments are now in order for what concerns the regime of validity of our results. In Sec.~\ref{sec:methods} we stated that the contribution to the chameleon effective mass from the photon coupling could be ignored. This amounts to the assumption that $\beta_m \rho_m \gg \beta_\gamma B^2/2$, which has been assumed in all subsequent calculations. Obviously, for sufficiently large values of $\beta_\gamma$ and in the presence of a sufficiently large background magnetic field this assumption may be spoiled: should this occur, we would need to make the replacement $\beta_m \rho_m \to \beta_m \rho_m + \beta_\gamma B^2/2$ in Eq.~\eqref{eq:m_eff}. We see that for the solar magnetic field model adopted, this would lead to a suppression in the solar flux for values of $\beta_\gamma \gtrsim 10^{15}$, given that the contribution from $\beta_\gamma B^2/2$ would lead to $m_{\text{eff}}^2>\omega^2$ in the solar regions with the strongest magnetic field. However, our model assumes that certain regions do not feature bulk magnetic fields -- therein obviously $\beta_\gamma B^2/2 = 0$, implying that the form of $m_\mathrm{eff}$ we have adopted is valid. Adopting only contributions to the flux from solar regions with $B=0$ according to our model, we find the upper limit $\beta_\gamma \lesssim 4\times10^{13} \ll 10^{15}$, implying that for our magnetic field model the effective chameleon mass contribution from the photon coupling can safely be ignored.

\change{By assuming a static solar and magnetic field model we have ignored the effects of solar dynamics, which is expected to be important in the outer convective zone. However the flux of chameleons produced in this region is too low to have an impact on the solar energy loss bound, or to be detected by current or near-future technologies. As the production rate is very low compared to that in the core this low-energy contribution is not displayed in Fig.~\ref{fig:primakoff-LT}.
Furthermore, a preliminary analysis suggests that at these low energies the production mechanisms discussed in this paper are subdominant to plasmon processes. As a result, we see the static solar model as a justifiable assumption.}

\begin{figure}
    \centering
    \includegraphics[width=\linewidth,keepaspectratio]{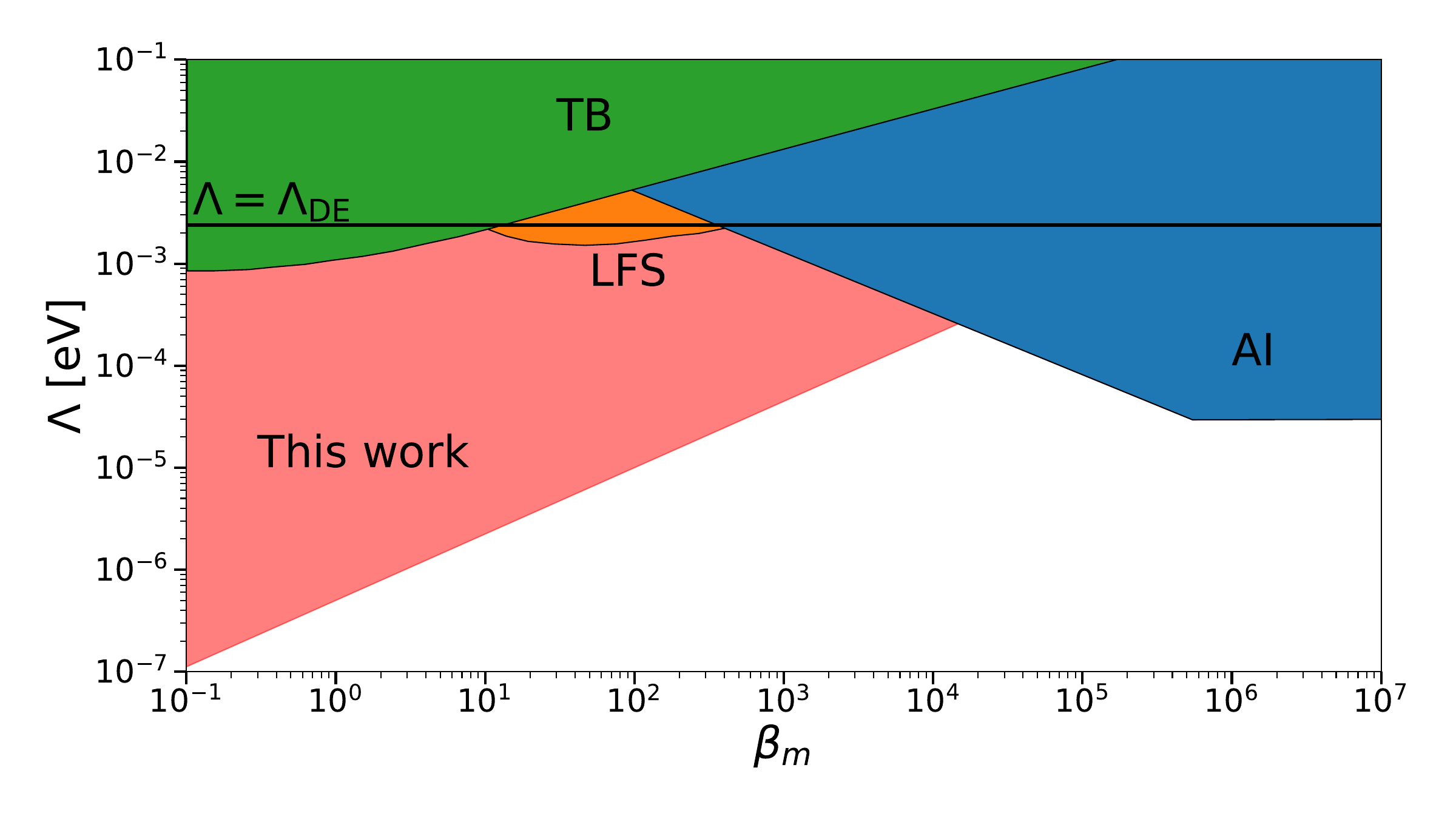}
    \caption{Comparison in the $\beta_m$-$\Lambda$ plane of the region of validity of the solar energy loss bound calculated in this paper against other existing experimental bounds, for $n=1$ chameleons (with the dark energy scale $\Lambda_{\text{DE}}=2.4\,{\text{meV}}$ marked by the black horizontal line). The shaded regions labeled ``TB'' (torsion balance, green region), ``AI'' (atom interferometry, blue region), and ``LFS'' (levitated force sensor, orange region) are excluded by the corresponding probes. Note that these previous bounds are independent of $\beta_\gamma$ and are therefore hard bounds, whereas the red shaded region labeled ``this work'' corresponds to the region for which the limit $\beta_\gamma < 10^{10}$ we derived in Eq.~\eqref{eq:solarluminosity} holds, as discussed in the main text. Solar chameleons, and in particular next-generation helioscopes, can therefore be used to explore a region of parameter space that has yet to be studied.}
    \label{fig:Lambda--reduced}
\end{figure}

Most of the observables relevant for fifth force experiments are independent of $\beta_{\gamma}$, and as a result these experiments can be used to place limits in the the $\beta_m$-$\Lambda$ parameter space that are independent of $\beta_{\gamma}$. On the other hand, the solar energy loss arguments discussed earlier, and based on the novel production mechanisms we have studied, set an upper limit on $\beta_\gamma$ for given values of $\Lambda$, $\beta_m$, and $n$. As such, unless we were to perform a global parameter scan (which is beyond the scope of the present work, but could be explored in a follow-up work), our analysis cannot place exclusion limits in the $\beta_m$-$\Lambda$ plane. Nevertheless, we can still identify the region for which the limit we obtained in Eq.~\eqref{eq:solarluminosity} holds, i.e.\ the region for which $m_{\text{eff}}^2\ll\omega^2$, so that the solar chameleon flux is completely independent of the chameleon model parameters other than $\beta_{\gamma}$ (even when considering values of $\Lambda$ other than the dark energy scale). This region is given by the red-shaded area of Fig.~\ref{fig:Lambda--reduced}. This section of parameter space can be interpreted as the region of parameter space of interest for future experiments that will be sensitive to solar chameleons, including next-generation helioscopes such as the International Axion Observatory (IAXO)~\cite{Armengaud:2014gea,IAXO:2019mpb} as well as its predecessor BabyIAXO~\cite{IAXO:2020wwp}. Such experiments will in fact be able to search for solar chameleons within the parameter space shown in Fig.~\ref{fig:Lambda--reduced} and, should no signal be observed, they will be able to place an upper limit on $\beta_\gamma$ which is more stringent than the solar energy loss limit calculated in this work. These and related aspects, including detailed forecasts for the future reach of IAXO and BabyIAXO, will be explored in a future work currently in progress.

As discussed in Sec.~\ref{sec:primakoff}, the $\phi F^2$ scalar-photon coupling results in a term proportional to $\phi (\mathbf{E}\cdot\mathbf{E} + \mathbf{B}\cdot\mathbf{B})$. This results in a difference in the available production mechanisms for scalars when compared with axions with their $a F \tilde{F} \propto a \mathbf{E}\cdot\mathbf{B}$ pseudoscalar coupling. The result is that axions can be produced by longitudinal plasmons (which have only an electric and no magnetic component) in the presence of an external magnetic field, and scalars in an electric field. The scalar-photon coupling also implies the existence of plasmon coalescence and decay processes, however it can be shown that these can be treated as limiting cases of the Primakoff process. Preliminary results for the longitudinal contributions to solar scalar production show that they are important at low energies and appear to dominate over the magnetic field production. This will be explored in detail in part II of our analysis, to be presented as a separate paper.

\section{Conclusions}
\label{sec:conclusions}

Motivated by the steadily growing interest in the search for new light particles, which could also be at the origin of dark matter and dark energy, we have reconsidered the possibility of producing (scalar) chameleon particles in the Sun. Chameleons are a well-motivated class of particles which, as a result of direct couplings to photons and matter fields, feature a density-dependent effective mass: this allows them to satisfy constraints from fifth-force searches on local scales, while remaining extremely light on cosmological scales and thereby potentially playing the role of dark energy and driving cosmic acceleration. Earlier studies considered the possibility of producing chameleons in the dense, hot (and magnetic) solar interior, but only considered production from the magnetic field in the tachocline~\cite{Brax:2010xq,Brax:2011wp,Vagnozzi:2021quy}. In the present work, we revisit this problem incorporating several previously overlooked details and production channels. Specifically, we include the contribution of the bulk magnetic field throughout the solar interior (thereby not limiting ourselves to the tachocline), as well as Primakoff production of chameleons in the electric fields of electrons and ions. Our results show that all these channels contribute significantly to the production rate and resulting flux of solar chameleons, which therefore has been completely revised compared to earlier studies. Finally, we have argued that the processes of photon coalescence and plasmon decay do not play a significant role in our derivation. We note that our results can be applied to both fixed-mass scalars and chameleons with density-dependent effective mass.

We have then used our revised solar chameleon flux to derive a novel bound on the chameleon-photon conformal coupling strength $\beta_{\gamma}$ by demanding that the total luminosity carried by solar chameleons does not exceed 3\% of the solar luminosity, as required by global fits to helioseismic and solar neutrino observables. We derive the upper limit $\beta_{\gamma} \lesssim 10^{10}$, a bound which is independent of other couplings to matter (and in particular of the conformal chameleon-matter coupling strength $\beta_m$). This constraint is shown in Fig.~\ref{fig:luminosity} along with the bound on $\beta_{\gamma}$ from CAST~\cite{CAST:2018bce}, as well as the most updated constraints on $\beta_m$, for both $n=1$ and $n=4$ chameleons (left and right panels respectively). We note, however, that the bound $\beta_{\gamma} \lesssim 5.7 \times 10^{10}$ from CAST has been derived assuming only production from the magnetic field in the tachocline. This limit clearly needs to be revised in light of our more complete derivation of the solar chameleon flux: since the previously computed flux actually represents a lower limit to the one we have derived, we expect that adopting our findings should tighten the CAST upper limit on $\beta_{\gamma}$, given that the solar chameleon flux scales as $\beta_{\gamma}^2$. Although the majority of our results have been derived setting the height of the chameleon potential $\Lambda$ to the dark energy scale $\Lambda_{\text{DE}}=2.4\,{\text{meV}}$, we have then extended our analysis considering more general values of $\Lambda$. The red shaded region in Fig.~\ref{fig:Lambda--reduced} shows the region of $\beta_m$-$\Lambda$ parameter space where our solar energy loss limit on $\beta_{\gamma}$ holds. This is compared to a number of other existing laboratory constraints, from atom interferometry, levitated force sensors, and torsion balance experiments. We clearly see that solar chameleons allow for the exploration of a region of chameleon parameter space that has yet to be covered.

Our study opens up various interesting avenues for potential follow-up work. The most pressing one is without doubt a dedicated study of the contribution of longitudinal plasmons to the solar chameleon flux, expected to be important especially at low energies. This will be studied in a separate work (part II of this series). With these results at hand, a dedicated study on the detectability of solar chameleons is then in order. In fact, the solar chameleons whose flux we have characterized can reach Earth and interact with terrestrial detectors. These include obviously next-generation axion helioscopes such as IAXO and BabyIAXO, as well as (multi-tonne) dark matter direct detection experiments. The present work is therefore naturally preparatory to studies on the detectability of solar chameleons in terrestrial experiments (in part updating the results of the earlier Ref.~\cite{Vagnozzi:2021quy}, which focused on dark matter direct detection experiments, in light of the updated solar chameleon flux). Such a possibility would naturally enhance the science reach of these experiments, primarily envisaged for the study of (among others) axions and weakly interacting massive particles, at zero extra cost. These and related possibilities are part of ongoing investigations which we plan to discuss in follow-up works which are currently in progress. Broadly speaking, it is our hope that our work will stimulate further studies assessing detection prospects of new light particles (including those equipped with screening mechanisms, not necessarily of the chameleon type). More importantly, considering the original theoretical motivation for chameleon particles, our work can also be the starting point for the exploration of non-gravitational interactions of dark energy, potentially leading us to unravel the physics behind cosmic acceleration in terrestrial laboratories and enabling direct detection of dark energy.

\vspace{0.5cm}

\begin{acknowledgments}
\noindent We thank Linda Raimondo for collaboration and discussions in all stages of the project. T.O'S., M.G., and J.K.V.\ acknowledge support from the Agencia Estatal de Investigaci{\'o}n (Spanish State Research Agency) through grant No.\ PID2019-108122GB-C31 funded by MCIN/AEI/10.13039/501100011033, from the ``European Union NextGenerationEU/PRTR'' (Planes complementarios, Programa de Astrof{\'i}sica y F{\'i}sica de Altas Energ{\'i}as), and from the European Union's Horizon 2020 research and innovation programme under the European Research Council (ERC) grant agreement ERC-2017-AdG788781. M.G.\ acknowledges support from the grant PGC2022-126078NB- C21 ``A\'{u}n m\'{a}s all\'{a} de los modelos est\'{a}ndar'' funded by MCIN/AEI/10.13039/501100011033 and ``ERDF A way of making Europe'', and from the DGA-FSE grant 2023-E21-23R funded by the Aragon Government and the European Union -- Next Generation EU Recovery and Resilience. A.C.D.\ is partially supported by the Science and Technology Facilities Council (STFC) through the STFC consolidated grant ST/T000694/1. S.V.\ acknowledges support from the University of Trento and the Provincia Autonoma di Trento (PAT, Autonomous Province of Trento) through the UniTrento Internal Call for Research 2023 grant ``Searching for Dark Energy off the beaten track'' (DARKTRACK, grant agreement No.\ E63C22000500003), and from the Istituto Nazionale di Fisica Nucleare (INFN) through the Commissione Scientifica Nazionale 4 (CSN4) Iniziativa Specifica ``Quantum Fields in Gravity, Cosmology and Black Holes'' (FLAG). L.V.\ acknowledges support by the National Natural Science Foundation of China (NSFC) through the grant No.\ 12350610240 ``Astrophysical Axion Laboratories'', as well as hospitality of the INFN divisions of Ferrara and Frascati National Laboratories and the Texas Center for Cosmology and Astroparticle Physics, Weinberg Institute for Theoretical Physics at the University of Texas at Austin (USA) during all stages of this work. This publication is based upon work from the COST Actions ``COSMIC WISPers'' (CA21106) and ``Addressing observational tensions in cosmology with systematics and fundamental physics (CosmoVerse)'' (CA21136), both supported by COST (European Cooperation in Science and Technology).
\end{acknowledgments}

\appendix
\section{Kinetic derivation of the conversion probability}
\label{sec:appendix}

The result derived in Sec.~\ref{sec:tft} can alternatively be obtained through a kinetic approach. For this, we recast the photon-scalar Lagrangian from Eq.~\eqref{eq:action} in the form:
\begin{equation}
     \label{eq:lagrangian-full}
    \begin{split}
     \mathcal{L} =&{} -\frac{1}{4}F_{\mu\nu}F^{\mu\nu} + \frac{1}{2}A_\mu \Pi^{\mu\nu}A_\nu + \frac{1}{2} \partial_\mu \phi \partial^\mu \phi \\
     &- \frac{1}{2} m^2 \phi^2 - \frac{\beta_\gamma}{4 M_\mathrm{Pl}} \phi F_{\mu\nu}F^{\mu\nu}\,,
    \end{split}
\end{equation}
where the polarization tensor $\Pi^{\mu\nu}$, which can be identified with the photon self-energy, results from approximating the induced current in a plasma as a linear response, and is given by the following:
\begin{equation}
    J^\nu = \frac{1}{2}\Pi^{\mu\nu} A_\mu\,.
\end{equation}
The mass term squared $m^2$ defined as in Eq.~\eqref{eq:m_eff} accounts for the effects of the self-interaction potential and the coupling of the chameleon to matter. The same result can be obtained including the full Lagrangian in Eq.~\eqref{eq:action1}. Eq.~\ref{eq:lagrangian-full} leads to the following equations of motion (EoM):
\begin{eqnarray}
    \left(\partial^2 + m^2\right) \phi &=& - \frac{\beta_\gamma}{4 M_\mathrm{Pl}} F_{\mu\nu}F^{\mu\nu}\,,
    \label{eq:EoM-phi1}\\
    \partial_\alpha F^{\alpha\beta} (1 \!+\! \frac{\beta_\gamma}{M_\mathrm{Pl}} \phi) &=& -\frac{\beta_\gamma}{M_\mathrm{Pl}} \partial_\alpha \phi F^{\alpha\beta} -A_\alpha \Pi^{\alpha\beta}\,.
    \label{eq:EoM-A1}
\end{eqnarray}
We are considering a constant external magnetic field, so treat the photon field $A^\mu$ as a background field plus a perturbation $A^\mu \to \bar A^\mu + \tilde A^\mu$, where the bar represents the background and the tilde the perturbation. We can do the same with the scalar field $\phi \to \bar \phi + \tilde \phi$ where the background field $\bar \phi$ is the field when only the external B-field is present, in which case Eq.~\eqref{eq:EoM-phi1} reduces to the following:
\begin{equation}
    0 = m^2 \bar \phi + \frac{\beta_\gamma B^2}{2 M_\mathrm{Pl}}\,,
\end{equation}
which in itself leads to the following field configuration:
\begin{equation}
    \bar \phi = \frac{-\beta_\gamma B^2}{2 M_\mathrm{Pl} m^2}\,.
    \label{eq:phibar}
\end{equation}
Eq.~\eqref{eq:EoM-phi1} with the perturbation can now be written as follows:
\begin{equation}
    \partial^2 \tilde \phi = -m^2 \tilde \phi - \frac{\beta_\gamma}{M_\mathrm{Pl}} \partial_\mu \tilde A_\nu \bar F^{\mu\nu}\,,
    \label{eq:EoM-phi2}
\end{equation}
where the background terms have cancelled and we have ignored higher order perturbation terms. Following the same steps with Eq.~\eqref{eq:EoM-A1} and taking the Lorenz gauge we reach the following:
\begin{equation}
    \partial^2 \tilde A^\beta (1 + \frac{\beta_\gamma}{M_\mathrm{Pl}} \bar \phi) + \frac{\beta_\gamma}{M_\mathrm{Pl}} \partial_\alpha \tilde \phi \bar F^{\alpha\beta} = - \Pi^{\alpha\beta}(\bar A_\alpha + \tilde A_\alpha) \,,
    \label{eq:EoM-A2}
\end{equation}
In the presence of only background fields we see $\Pi^{\alpha\beta}\bar A_\alpha = 0$, and using $\bar \phi$ in Eq.~\eqref{eq:phibar} yields the following:
\begin{equation}
    \partial^2 \tilde A^\beta (1 - \frac{\beta_\gamma^2 B^2}{2 M_\mathrm{Pl}^2 m^2}) + \frac{\beta_\gamma}{M_\mathrm{Pl}} \partial_\alpha \tilde \phi \bar F^{\alpha\beta} = - \Pi^{\alpha\beta} \tilde A_\alpha \,,
    \label{eq:EoM-A3}
\end{equation}
and using the form of $\bar F^{\mu\nu}$ for an external B-field in the $z$-direction, we obtain the following EoM:
\begin{equation}
    ((1-a^2) \partial^2 + \pi_y)A_y - \frac{\beta_\gamma}{M_\mathrm{Pl}} \partial_x \phi B_z = 0\,,
    \label{eq:EoM-A4}
\end{equation}
\begin{equation}
    (\partial^2 + m^2)\phi + \frac{\beta_\gamma}{M_\mathrm{Pl}} \partial_x A_y B_z = 0\,,
    \label{eq:EoM-phi3}
\end{equation}
where $a^2\equiv \frac{\beta_\gamma^2 B^2}{2 M_\mathrm{Pl}^2 m^2}$, and we have dropped the tildes. After Fourier transforming and performing the field redefinitions $A \to \sqrt{1-a^2} A$, $\beta_\gamma \to \beta_\gamma/\sqrt{1-a^2}$, and $\pi_y \to \pi_y/\sqrt{1-a^2}$ we get the EoM:
\begin{equation}
    \left( \omega^2 - k^2 - \left( \begin{array}{cc}
         \pi_y & i \chi \\
         -i \chi & m^2
    \end{array} \right) \right) \left( \begin{array}{c}
         A_y \\
         \phi
    \end{array} \right) = 0\,,
    \label{eq:EoM-matrix1}
\end{equation}
where $\chi \equiv \beta_\gamma k_x B_z / M_\mathrm{Pl}$. We can diagonalize this rotating by the following matrix:
\begin{equation}
    R \equiv \left( \begin{array}{cc}
         \cos\theta & i \sin\theta \\
         i \sin\theta & \cos\theta
    \end{array} \right)
\end{equation}
whose inverse is just its Hermitian conjugate, with the rotation angle given by:
\begin{equation}
    \tan 2\theta = \frac{2\chi}{m^2 - \pi_y}\,.
\end{equation}
This results in the following EoM:
\begin{equation}
    \left[\! \omega^2 \!-\! k^2 \!-\! \left(\!
    \begin{array}{cc}
         \pi_y + \mathcal{O}(M_\mathrm{Pl}^{-2}) & 0 \\
         0 & m^2 + \mathcal{O}(M_\mathrm{Pl}^{-2})
    \end{array}
    \! \right)\! \right]\!\left(\!
    \begin{array}{c}
         A_y' \\
         \phi'
    \end{array} \!\right) \!=\! 0\,.
    \label{eq:EoM-matrix2}
\end{equation}
As these primed states propagate freely, a formal solution for their evolution is as follows:
\begin{equation}
    \left( \begin{array}{c}
        A'(L)\\
        \phi'(L)
    \end{array} \right) = \left( \begin{array}{cc}
        e^{-i k_\gamma L} & 0 \\
        0 & e^{-i k_\phi L}
    \end{array} \right) \left( \begin{array}{c}
        A' \\
        \phi'
    \end{array} \right) \,,
    \label{eq:propogation}
\end{equation}
so that the original states can be obtained by rotating back into the original basis via a matrix describing the mixing of the different states:
\begin{equation}
    \left( \begin{array}{c}
        A_y(L)\\
        \phi(L)
    \end{array} \right) = R^{-1} \left( \begin{array}{cc}
        e^{-i k_\gamma L} & 0 \\
        0 & e^{-i k_\phi L}
    \end{array} \right) R \left( \begin{array}{c}
        A_y \\
        \phi
    \end{array} \right)\,.
    \label{eq:propogation2}
\end{equation}
The conversion probability is given by the diagonal term squared, that is:
\begin{equation}
    P_{\gamma \to \phi}(L) = \left| \frac{\sin 2\theta}{2} (e^{-i k_\phi L} - e^{-i k_\gamma L} ) \right|^2\,.
    \label{eq:conversionprob}
\end{equation}
Assuming $\theta \ll 1$, which corresponds to the weak coupling we expect, we can use the fact that $\sin2\theta \approx \tan2\theta$. We can see that $k_\phi = \sqrt{\omega^2 - m^2}$, and using the following result from Weldon~\cite{Weldon:1983jn}:
\begin{equation}
    \pi_y = m_\gamma^2 - i \omega \Gamma_\gamma
    \label{eq:pi_y}
\end{equation}
we find $k_\gamma \approx \sqrt{\omega^2 - m_\gamma^2} + i \Gamma_\gamma / 2$ for $\Gamma_\gamma \ll m_\gamma$. We have assumed the following form:
\begin{equation}
    \Gamma_\gamma \!=\! \frac{64\pi^2\alpha^3}{3m_e^2\omega^3}
    \sqrt{\frac{m_e}{2\pi T}} n_e \left( 1\!-\!e^{-\omega/T} \right)
    \sum_i Z^2 n_i F_i
    + \frac{8\pi\alpha^2}{3m_e^2} n_e\,,
    \label{eq:Gamma_gamma}
\end{equation}
where the sum is over all ion species in the Sun, and $F_i$ is the thermally-averaged Gaunt factor for species $i$. For the numerical results $F_i$ was taken from Ref.~\cite{Chluba:2019ser} and the sum over ions has been assumed to include only hydrogen and helium. The conversion probability can finally be written in the following form:
\begin{equation}
    \label{eq:conversionprob2}
    \begin{split}
    P_{\gamma \to \phi}(L) \approx&{} \left( \frac{\beta_\gamma B_z}{M_\mathrm{Pl} \sqrt{1-a^2}} \right)^2 \frac{\omega^2}{(m^2 - m_\gamma^2)^2 + (\omega\Gamma)^2}\\
    &\times \left( 1 + e^{-\Gamma L} - 2 e^{-\Gamma L / 2} \cos(qL) \right)\,,
    \end{split}
\end{equation}
where $q \equiv \sqrt{\omega^2 - m_\gamma^2} - \sqrt{\omega^2 - m^2}$. The factor $\sqrt{1-a^2}$ from the redefinition of $\beta_\gamma$ has been shown explicitly here, but we will assume from now that $a^2 \ll 1$ so $\sqrt{1-a^2}\to1$. For production in the solar interior, for $\phi$ to escape the Sun $\Gamma L \gg 1$ is required, so the conversion probability reduces to the the following $L$-independent expression:
\begin{equation}
     P_{\gamma \to \phi} \to \left( \frac{\beta_\gamma B_z}{M_\mathrm{Pl}} \right)^2 \frac{k_\phi^2}{(m^2 - m_\gamma^2)^2 + (\omega\Gamma)^2} \,.
\end{equation}
We can then use Eq.~\eqref{eq:dN} as in Sec.~\ref{sec:tft}, with the following definition:
\begin{equation}
    \Gamma_\phi^{\mathrm{pro}} = \frac{\Gamma_\gamma P_{\gamma \to \phi}}{e^{\omega/T} - 1}\,,
\end{equation}
with production rate then given by:
\begin{equation}
    \label{eq:dN--kinetic}
    \frac{\mathrm{d}\dot N}{\mathrm{d}\omega} \!=\! \frac{2 \beta_\gamma^2}{\pi M_\mathrm{Pl}^2} \!\int_0^{R_\odot} \!\!\! r^2 \mathrm{d}r B_z^2(r) \frac{\omega(\omega^2 - m^2)^{3/2}}{(m_\gamma^2 - m^2)^2 \!+\! (\omega \Gamma_\gamma)^2} \frac{\Gamma_\gamma}{e^{\omega/T} \!-\! 1}.
\end{equation}
Identifying $B_z$ with $B_\perp$, we see that Eq.~\eqref{eq:dN--kinetic} is identical to Eq.~\eqref{eq:dN_TFT--2}.

\bibliographystyle{apsrev4-1}
\bibliography{solarchameleon.bib}

\end{document}